\documentclass[runningheads]{llncs}
\usepackage[T1]{fontenc}
\usepackage{graphicx}
\usepackage{enumitem}
\usepackage{mathtools}
\begin{document}
\title{Divide and Conquer based Symbolic Vulnerability Detection}
%
%
\author{Christopher Scherb\inst{1}\orcidID{0000-0001-6116-5093} \and
Luc Bryan Heitz\inst{1}\orcidID{0009-0000-9279-2751} \and
Hermann Grieder\inst{1}\orcidID{0000-0001-9984-8615}}
\authorrunning{C. Scherb et al.}
%
\institute{University of Applied Sciences and Arts Northwestern Switzerland
\email{christopher.scherb@fhnw.ch, luc.heitz@fhnw.ch, hermann.grieder@fhnw.ch}
}
\maketitle              
\begin{abstract}
In modern software development, vulnerability detection is crucial due to the inevitability of bugs and vulnerabilities in complex software systems. Effective detection and elimination of these vulnerabilities during the testing phase are essential. Current methods, such as fuzzing, are widely used for this purpose. While fuzzing is efficient in identifying a broad range of bugs and vulnerabilities by using random mutations or generations, it does not guarantee correctness or absence of vulnerabilities. Therefore, non-random methods are preferable for ensuring the safety and security of critical infrastructure and control systems. This paper presents a vulnerability detection approach based on symbolic execution and control flow graph analysis to identify various types of software weaknesses. Our approach employs a divide-and-conquer algorithm to eliminate irrelevant program information, thus accelerating the process and enabling the analysis of larger programs compared to traditional symbolic execution and model checking methods.

\end{abstract}
\section{Introduction}
Bug and vulnerability detection is a topic as old as software development itself. When developing complex and sophisticated software bugs naturally occur. Since the beginning of software development, the goal of engineers and researchers always was to achieve the 
maximum of what was feasible with the current technology. While the technology has advanced and even problems which were hard to solve in the 1990s and 2000s are now easily solvable, hand in hand with the advancement of technology, and thus the complexity of the problems also increased. Therefore, often we develop technology on the edge what is possible. Pushing the limit of what is possible also can lead to mistakes which result in bugs and vulnerabilities. 
Therefore, for software which should be used in productive environments, it is very advisable to have a sophisticated testing chain to reduce the number and the impact of possible mistakes. 
In this paper we present a test method based on symbolic execution, which aims to reliably find certain types of bugs and vulnerabilities which are largely based on memory corruption, but not exclusively. Our approach focuses on software for devices such as Operational Technology (OT) and the Internet of Things (IoT), where software is relatively small but often hard to update. 

\subsection{Testing}

Over time different test methods emerged, starting with unit tests, integration tests and also vulnerability testing. Testing for bugs and vulnerabilities is not always trivial, since most unit and integration tests are focused on ensuring the functionality works as intended. However, bugs and vulnerabilities primarily occur due to insufficient fail-safe mechanisms, especially in input data parser or interpreters. Finding all possible failures and exceptions to create a fail-safe for any case is a complicated and complex task, as there are almost infinite possibilities how something can go wrong, however only one case how everything works correctly. 

To address this problem, testing for bugs and vulnerability significantly differs from unit and integration tests. First and foremost to develop secure software, a secure design is required. However, implementation -- even if the design is secure -- still can introduce bugs and vulnerabilities. To search for these vulnerabilities typically different testing methods are applied, such as static analysis~\cite{zhioua2014static}, Software Composition Analysis (SCA)~\cite{imtiaz2021comparative} as well as dynamic analysis/fuzzing~\cite{zhu2022fuzzing}\cite{serebryany2012addresssanitizer}. Furthermore, manual code reviews and threat modeling are typical approaches to extend the automatized testing methods. 

Generally, all these methods are well working within their limitations. 
Static analysis is a very robust method, which can detect many bugs, however, it is limited to analysis of the code itself, as it does not execute the program and, thus, lacks runtime context. Therefore, it cannot detect bugs such as e.g., race conditions, memory leaks and use-after-free vulnerabilities. 
SCA has other limitations as it uses databases to detect vulnerable components and cannot find new bugs and vulnerabilities on its own. Lastly, Dynamic analysis and fuzzing can find new bugs but usually are not able to cover the entire program.
Fuzzing with many extensions has proven itself to be one of the most efficient techniques to find even complex bugs and vulnerabilities. However, fuzzing is based on randomized inputs and thus cannot guarantee the absence of bugs. 

\subsection{Verification}

The ideal case of software testing would be the verification, that is that the software is correct and contains no bugs. However, this kind of verification is very difficult since correctness is not defined on the machine level but by the specifications. Nevertheless, 
it is possible to model software and to verify the model in regard to defined specifications~\cite{chan1998model}. The limitation of these checks are of course given by how precise the created model represents the actual software. 
Verification of software is a very labor intensive process and in reality it is barely used. 

Other methods that can reach a certain level of verification, such as symbolic execution~\cite{baldoni2018survey}\cite{king1976symbolic} do not require a model to be created, however, they are not designed for vulnerability detection but for reasoning about the program behavior. However, by simulating the entire software, symbolic execution is very slow, even though there are some approaches to reduce the problem size.

\subsection{Contribution \& Limitations}
This paper suggests a reduced symbolic execution approach which is focused only on the detection of certain vulnerabilities. Whereby, in a first step, programs are sliced into ranges~\cite{haltermann2024ranged} or functions~\cite{anand2008demand} and the individual parts analyzed for vulnerabilities. Meanwhile, from each part features relevant to vulnerability detection are extracted and later analyzed for more-complex-to-detect vulnerabilities. 
By splitting up the computational expensive symbolic execution into smaller parts we dramatically decrease the execution time. Moreover, stripping away all features and side effects which are not relevant to vulnerability detection further decreases the execution time. 

However, while our approach shows some performance gains in vulnerability detection, we are still limited to relatively small problems. Nevertheless, our approach makes it feasible to verify the absence of certain vulnerabilities in smaller pieces of software, such as in microcontroller, industrial control systems, embedded medical devices, etc~\cite{ani2017review}. 

This field of devices is critical in regard of cybersecurity because often no update mechanisms are available and code needs to be certified before updates can be released. Thus, removing as many vulnerabilities as possible during the development process here is more crucial compared to other areas of software development.  

Our \emph{main contribution} in this paper is an algorithm to extract security-relevant features using symbolic execution from individual parts of a computer program, and later recombine only these features for a reliable and faster detection of these vulnerabilities. If we cannot find a certain type of bug in a program, it proved the absence of this type of bug. 

\subsection{Structure}

Our paper is structured as follows: We start by giving an overview over the related work (Section \ref{sec:relatedwork}). 
Afterwards we present our approach of a divide and conquer strategy for symbolic vulnerability detection in multiple steps: First we give an overview over the interaction of the different components (Section \ref{sec:overview}), before we dive into details of each individual phase: divide (Section \ref{sec:divide}), conquer (Section \ref{sec:conquer}) and  weakness modeling (Section \ref{sec:vul}).
Before we conclude our paper we will present our evaluation (Section \ref{sec:eval}).

\section{Background \& Related Work}
\label{sec:relatedwork}

Generally, there are several approaches to make symbolic execution more usable in practice, ranging from fuzzing extensions to help fuzzers~\cite{chen2018systematic} to overcome magic bytes to strategies to split programs into individual parts and recombine them. 

\subsection{Symbolic Execution}

Symbolic execution is a program analysis technique that is used to systematically explore program paths by treating input values as symbolic variables rather than concrete values. This approach enables the analysis of multiple execution paths simultaneously, allowing for the detection of bugs, generation of test cases, and verification of program properties. By using symbolic constraints to represent program inputs, symbolic execution can uncover errors and vulnerabilities that might be missed by traditional testing methods. However, symbolic execution suffers from a path explosion problem, since the number of possible execution paths grows exponentially with the size of the computer program. Moreover, only the constraint solving required by symbolic execution is already at least NP-complete~\cite{traxler2008time}.

Therefore, many newer approaches to symbolic execution try to reduce the number of branches or the number of constraints by pruning them or by reducing the problem size~\cite{yi2017eliminating}. 

\emph{Compositional Dynamic Test Generation} is a technique developed by Godefroid~\cite{godefroid2007compositional} which decomposes a computer program into components (for example into individual functions) and executes each component individually. Since a component is far smaller than an entire computer program, it is possible to symbolically execute them and to create summaries of the individual program components. Later the summaries are recombined to gain an overall understanding of the entire program. More fine graduated the summaries lead to a better performance gain~\cite{lin2015compositional}. 

\emph{Chopped symbolic execution} is a method that aims to address the path explosion problem and increase the efficiency of traditional symbolic execution. This method introduces a chopping criterion to symbolically execute only a relevant part of the program, thereby ignoring parts that do not affect the outcome of interest. The criterion is based on a user-specified property of the code or a set of variables the user is interested in~\cite{trabish2018chopped}.

\emph{Loops} are a big problem for symbolic execution. There are different approaches to improve loop handling either by creating loop summaries, such as done by Java Pathfinder~\cite{havelund2000model} or KLEE~\cite{xie2015s}. Other approaches try to find loop invariants using symbolic execution~\cite{sharma2012interpolants} or limit the number of iterations which are analyzed~\cite{clarke2004tool}. However, this technique may miss behaviors that occur beyond the set limit. 

\subsection{Symbolic Execution and Vulnerability Detection}

Symbolic execution has been widely used for vulnerability detection and software testing. However, as program complexity grows, the path explosion problem becomes a significant challenge. Recent research has focused on various techniques to address this issue and improve the efficiency of symbolic execution for vulnerability detection.
Lin et al.~\cite{li2013software} propose Fine-Grained Summaries, which create summaries for small code units such as individual lines or small blocks, rather than entire functions. This finer granularity aims to provide more precise representations of program behavior, potentially leading to more accurate analysis and efficient path exploration.
Yao et al.~\cite{yao2017statsym} present StatSym, a framework that combines statistical analysis with symbolic execution. StatSym constructs predicates and candidate paths likely to contain vulnerabilities using runtime information. These are then used to guide symbolic execution, effectively pruning the search space and potentially accelerating vulnerability detection.
Luckow et al.~\cite{luckow2020complexity} focus on detecting worst-case complexity vulnerabilities. Their technique uses context-preserving histories to guide symbolic execution towards paths likely to exhibit worst-case behavior. By learning policies from small input sizes and applying them to larger inputs, this approach aims to improve scalability in complexity analysis.
In the domain of hardware security, Tang et al.~\cite{tang2022accelerating} propose a method for detecting vulnerabilities in System on Chip (SoC) designs. Their approach involves converting hardware designs to C++ code and then using the KLEE symbolic execution engine to perform state exploration. By employing heuristic search strategies, they aim to accelerate the state space search and pinpoint security vulnerabilities more efficiently.
Tu et al.~\cite{tu2023boosting} propose a technique to boost symbolic execution for heap-based vulnerability detection and exploit generation. Their approach combines a new path exploration strategy, a novel memory model, and a new environment modeling solution to improve the efficiency and effectiveness of symbolic execution in detecting and exploiting heap-based vulnerabilities.

Since symbolic execution alone does not scale to large computer programs, other means of testing have been applied in combination with symbolic execution. For example driller~\cite{stephens2016driller} and SymCC~\cite{poeplau2020symbolic} combines fuzzing with symbolic execution, where the fuzzing process is mainly used for vulnerability detection, the symbolic execution is used to overcome magic bytes and other constraints which are hard to be handled by a fuzzer alone. 

KLEE and S2E~\cite{chipounov2011s2e} have been shown to be very efficient in path exploration for vulnerability detection. They aim to generate inputs that may create issues that are undetectable by pure static analysis. Typically, symbolic execution can check for vulnerability conditions such as an \emph{symbolic instruction pointer} (a memory corruption bug if the instruction pointer depends on the user input) or unsafe functions which depend on user input~\cite{ramos2015under}. 

Taint analysis has become a significant technique in the field of vulnerability detection, focusing on tracking the flow of sensitive information within software systems to identify security flaws. Taint analysis works by marking data from untrusted sources as "tainted" and monitoring its propagation through the program to detect potential misuse that could lead to vulnerabilities such as SQL injection, Cross-Site Scripting (XSS), and buffer overflows.

Dynamic taint analysis~\cite{newsome2005dynamic} involves analyzing a program execution to track the flow of tainted data in real time, which makes it effective for detecting runtime vulnerabilities that arise from user input. For example, the combination of concolic execution with dynamic taint analysis has shown promise in identifying XSS vulnerabilities by tracking tainted variables and their propagation paths during program execution~\cite{s23239407}.
Dynamic taint analysis can be powerful to detect certain vulnerabilities, especially in combination with symbolic execution~\cite{schwartz2010all}.

\subsection{Vulnerabilities and Weaknesses}
In cybersecurity, the terms "weakness" and "vulnerability" are often used interchangeably but denote different concepts. A weakness refers to a flaw or a shortcoming in the design, implementation, or configuration of a system that could potentially be exploited to cause harm~\cite{Shostack2014}. Common Weakness Enumeration (CWE) is a structured classification of software weaknesses maintained by the MITRE Corporation, providing a standardized language to describe these flaws~\cite{MITRECWE}. On the other hand, a vulnerability is a specific instance of a weakness that can be exploited by a threat actor to perform unauthorized actions within a system~\cite{NIST2012}. Vulnerabilities can lead to significant security breaches, as evidenced by numerous high-profile incidents in recent years. The National Institute of Standards and Technology (NIST) defines vulnerability as a flaw or weakness in the design, implementation, operation, or management of a system that could be exploited to violate the security policy of the system~\cite{NIST800-30}. Understanding the distinctions and relationships between weaknesses and vulnerabilities is crucial for developing robust security measures and mitigating potential risks in information systems.

\subsection{Differentiation from Previous Work}
Our approach to symbolic execution and vulnerability detection introduces several key innovations. Unlike the fine-grained summaries of Lin et al. [14] or function-level summaries of Godefroid [10], we create security-focused summaries that capture only vulnerability-relevant information. This results in more compact and targeted analysis.
We maintain a high-level view of the program structure, differing from StatSym's [1] runtime statistical analysis or Luckow et al.'s [2] context-preserving histories. This enables efficient combination of summaries across different program parts, allowing for comprehensive vulnerability analysis.
Our method scales effectively to large, complex programs, unlike Tang et al.'s [3] SoC-specific approach or Tu et al.'s [4] heap-focused method. By concentrating on security-relevant information and employing efficient summary combination, we can analyze extensive codebases more thoroughly.
We specifically target complexity vulnerabilities, using our focused summaries to identify paths leading to algorithmic complexity issues. This differs from Luckow et al.'s [2] approach in how we identify and analyze these vulnerabilities.
Finally, our reliance on static analysis distinguishes us from approaches requiring runtime information or environment modeling. This makes our technique applicable to a broader range of scenarios, including situations where runtime data is unavailable.
By integrating these aspects, our work advances symbolic execution for complexity vulnerability detection, offering a more scalable, focused, and widely applicable approach for large, real-world programs.

In this paper our focus is on memory corruption vulnerabilities~\cite{van2012memory} where either out-of-bounds read or writes occur, stack or heap metadata are corrupted, or non-allocated memory or freed memory is used or reused.

\section{Overview: Divide and Conquer based Symbolic Vulnerability Detection}
\label{sec:overview}

\begin{figure}
    \centering
    \includegraphics[width=0.5\textwidth]{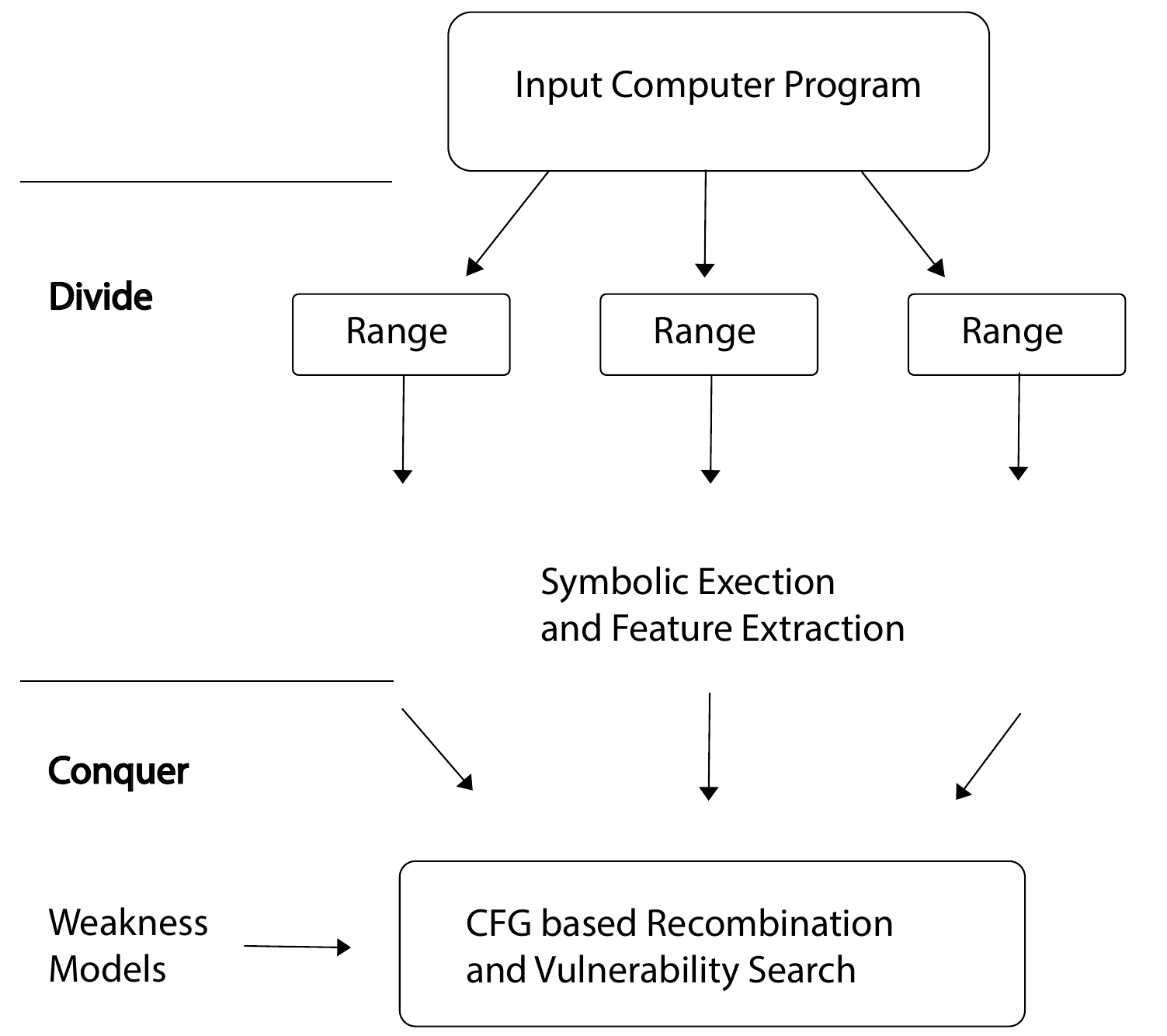}
    \caption{Overview of the Divide and Conquer Algorithm for Symbolic Vulnerability Detection}
    \label{fig:se_overview}
\end{figure}

The main idea of our approach to speed up the vulnerability detection process using symbolic execution is to split the program into smaller parts, which can be symbolically executed. Generally, symbolic execution can reason about program behavior and has many use-cases such as program analysis, debugging, software verification, optimization, automated test case generation, vulnerability detection, and many more. Most approaches to improve the speed of symbolic execution focus on preserving the full range of features. 
For us, the main focus is on vulnerability detection, thus our approach does not aim to preserve the full range of features, but is designed to be fast in vulnerability detection. Figure \ref{fig:se_overview} gives an overview of the process.

\paragraph{Divide.} We start by splitting up the program to be tested into different slices. These slices can be defined by functions or any other smaller or larger slice in the the computer program. The individual ranges of the computer program are executed and checked for vulnerabilities, such as buffer-overflows or other memory corruption issues, which can be detected within the range. Moreover, we extract features from each range. These features are used to create smart summaries of a range, which contain all information required to detect more complex vulnerabilities. Since the features are only containing information relevant for vulnerabilities, they are more compact than describing the entire behavior of a function.

\paragraph{Conquer.} In the \emph{conquer} phase we put the features we extracted during the \emph{divide} phase back together. Therefore, we use the Control Flow Graph (CFG)~\cite{beyer2011cpachecker} of the computer program, which defines all possible paths that the computer program can take. Instead of execution the entire path, as classic symbolic execution would do, we only execute the relevant parts which we need to know if certain vulnerabilities are there. For example, to detect the Use-After-Free Weakness, which is one of the most occurring once, we only need to execute malloc and free statements. 
If a possible vulnerability is detected, we use guided backward symbolic execution to verify that the taken path was actually possible in the real program. Only if a real world input exists that can trigger the vulnerability, we mark it as a vulnerability. 

\paragraph{Weakness Modeling.} To detect vulnerabilities we need to have a weakness model, that we can match by the divide and conquer algorithm. Simple vulnerabilities can be detected by checking if the instruction pointer became symbolic during the execution of a program or range. This means, that the instruction pointer was overwritten, for example by a stack-buffer-overflow or similar issues. More complex vulnerabilities such as Use-After-Free vulnerabilities may occur due to the combination of multiple functions or even threads and cannot be detected within a single range. Here the recombination during the conquer phase is important. Depending on the extracted features it is possible to detect a variety of weaknesses such as Use-After-Free, Duplicated-Free, Null-Pointer-De-References, Incorrect-Neutralization-Of-Special-Elements, etc. We modeled some weaknesses to detect corresponding vulnerabilities.

\section{Divide: Smart Symbolic Summaries}
\label{sec:divide}
\begin{figure}
    \centering
    \includegraphics[width=0.6\textwidth]{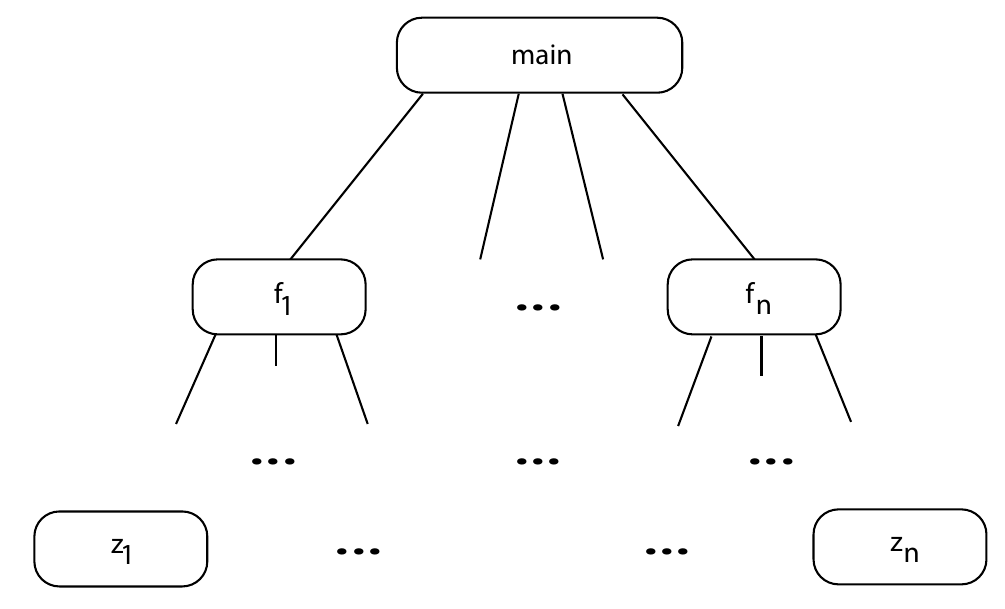}
    \caption{Function Call Graph of a Computer Program}
    \label{fig:functioncallgraph}
\end{figure}

The divide phase is the first phase of our divide and conquer algorithm. It aims to split the computer program into ranges. A range is defined as a reusable subpart of the computer program. In theory, any slice through the computer program can be a range, but in our case, it makes sense to focus on reusable parts, for example, methods/functions, since they are a natural unit of organization of the code. However, for very long functions it makes sense to introduce additional ranges. 
Our divide algorithm starts by splitting the computer program along function calls. The principle is quite straightforward. We generate a function call graph of the computer program, as shown in Figure \ref{fig:functioncallgraph}, and start executing the functions which do not have any dependencies to other functions, (at the bottom of the CFG, the functions $z_1$ to $z_n$ in the Figure). By executing and summarizing the inner functions first, we can use them to execute outer functions ($f_1$ to $f_n$ in the Figure) and so on, until we are up at the main function. 

\paragraph{Creating Symbolic Summary.} We define a \emph{symbolic summary} as a transformation of a symbolic input parameter into a symbolic result. A symbolic result is a constraint value that the function returns with a certain defined input parameter. The input parameter can be either a concrete or symbolic value.
Our summaries are based on lookup tables in a way that when an input parameter is given it is matched against the lookup table and the matching symbolic result is returned. 
We create a \emph{symbolic summary} by symbolically executing a function to find all possible symbolic results of the function. Next, we detect by reverse symbolic execution, which symbolic result is caused by which input parameter. Finally we store the mapping $m$ from input parameter to symbolic result:
\begin{equation}
    \label{equ:symmapping}
    m = (p_1, p_2, ... p_n) \rightarrow r_s,
\end{equation}
where $p_1, p_2, ... p_n$ are the constraint symbolic input parameter given to a function and $r_s$ is the constraint symbolic result the function will return given the parameter. 

The set of all possible results $r_s$ and the corresponding parameter is a function summary $S$ for a function $f$.

\begin{equation}
    \label{equ:functionsummary}
    S(f) = \{m_1, .... m_x\},
\end{equation}
where $x$ is the number of different paths through the function. 

For example, the mapping as shown in equation \ref{equ:symmapping} can be defined for the function $f$: 
\begin{verbatim}
            int f (int x, int y){
                if (x > 10){
                    return y 
                }
                else {
                    return x
                }
            }
\end{verbatim}
will be $(x > 10, y) \rightarrow y$ and $(x <= 10, y)  \rightarrow x$. Since in this example the return value depends only on the value of the input parameter $x$, $x$ is constrained in the summary, while $y$ is unconstrained. 
Since there are no more paths through the given function, the function summary can be written as: 

\begin{equation*}
    \label{equ:examplesummary}
    f(x,y) =  
\begin{cases}
 & \text{ if } x > 10 \rightarrow y \\
 & \text{ else } \rightarrow x 
\end{cases} 
\end{equation*}

When a function g, which calls f is executed, the summary is applied instead of executing f every time. By applying summaries, we widen the parameter space, 
which means, it becomes under-constrained. Therefore, the approach generally can produce false-positives, but not false-negatives. It may reach paths that are not reachable by the real program, but it will reach all paths which are reachable by the real program. 

\paragraph{Vulnerability Detection.} While executing a function to create a symbolic summary, we can already start searching for vulnerabilities. A typical starting point is searching for the input parameter of the function, which will create a situation where the instruction pointer becomes unconstrained. The instruction pointer typically points to the next instruction which is executed by a computer program. The instruction pointer usually moves to the next instruction after executing the current one or is moved to a predefined location by a \texttt{jmp} or \texttt{call} instruction. Even so, this location can be runtime defined (indirect \texttt{jmp}/\texttt{call}), and the locations are clearly defined.
However, a situation, where the next location of the instruction pointer solely depended on user input can only be created by a memory corruption vulnerability such as a buffer-overflow or a Use-After-Free . In such a case, the instruction pointer becomes unconstrained when the program is executed symbolically. Therefore, searching for unconstrained instruction pointers is a good way to look for vulnerabilities. Since our approach is under-constraint and may lead to false positives, we use guided backward symbolic execution to validate if it is possible to create an input triggering the potential vulnerability. 
Generally, many vulnerabilities in computer program can be caught by symbolically executing a computer program and searching for unconstrained instruction pointers. 
However, to find vulnerabilities caused by the order of multiple function calls -- such as Use-After-Frees -- in more complex computer program it is require to symbolically execute the entire computer program. Since this is often impossible due to the computational complexity~\cite{hensel2018termination}\cite{cook2013ranking}, we extract features during the execution of a function and later use these features to find vulnerable paths. 

\paragraph{Feature Extraction.}
Beside the initial vulnerability detection, we also extract features during the symbolic execution of the individual functions. The goal is to \emph{remember} operations performed by the function, which can have a security impact. This strongly depends on what vulnerabilities we want to search for. For example, when we search for Use-After-Free issues, we need to extract details about heap operations such as \texttt{malloc} and \texttt{free}, but when we search, for example, for SQL injections we need to extract details about database access and scan them for unsanitized inputs. This can, for example, be achieved by hooking the functions that access the database~\cite{malmain2024libafl}.

Thus, feature extraction is a process that depends a lot on the use-case. Therefore, we define the features we want to extract explicitly. A feature can be a function call, a library call, or any other symbolic condition. Moreover, a feature can be conditional or not. A conditional feature contains a condition on the input parameter, under which the feature is triggered. 
An unconditional feature can be written as: $f: malloc(type)>addr \rightarrow free(addr)$, which describes a function $f$ that calls $malloc$ and $free$ on the same memory address. 
A conditional feature can be written as $f: malloc(type)>addr$ \texttt{if x > 4}  $ \&\&  $ \texttt{a <100}, which describes a feature, where \texttt{malloc} is called if an input parameter $x$ is greater than $4$ and second input parameter $a$ is smaller than 100.

\section{Conquer: Summary Graph Analysis}
\label{sec:conquer}

The conquer phase aims to put everything back together. We use the CFG of the computer program to put the features we extracted back into the logical order and then we use graph analysis and model checking methods to search for vulnerabilities. The logical order is created by traversing the CFG in the first order of depth. In case paths get to long (for example due to recursive function calls), it has to be limited, comparable with loop limiters used by other symbolic execution approaches. 

When traversing the graph in depth-first order, we will keep track of the occurrence of the features. Whenever we pass by a feature, we verify the feature against weakness models. Some weakness models require us to track the state over, while others can be decided on the spot. 

\begin{figure}
    \centering
    \includegraphics[width=0.9\textwidth]{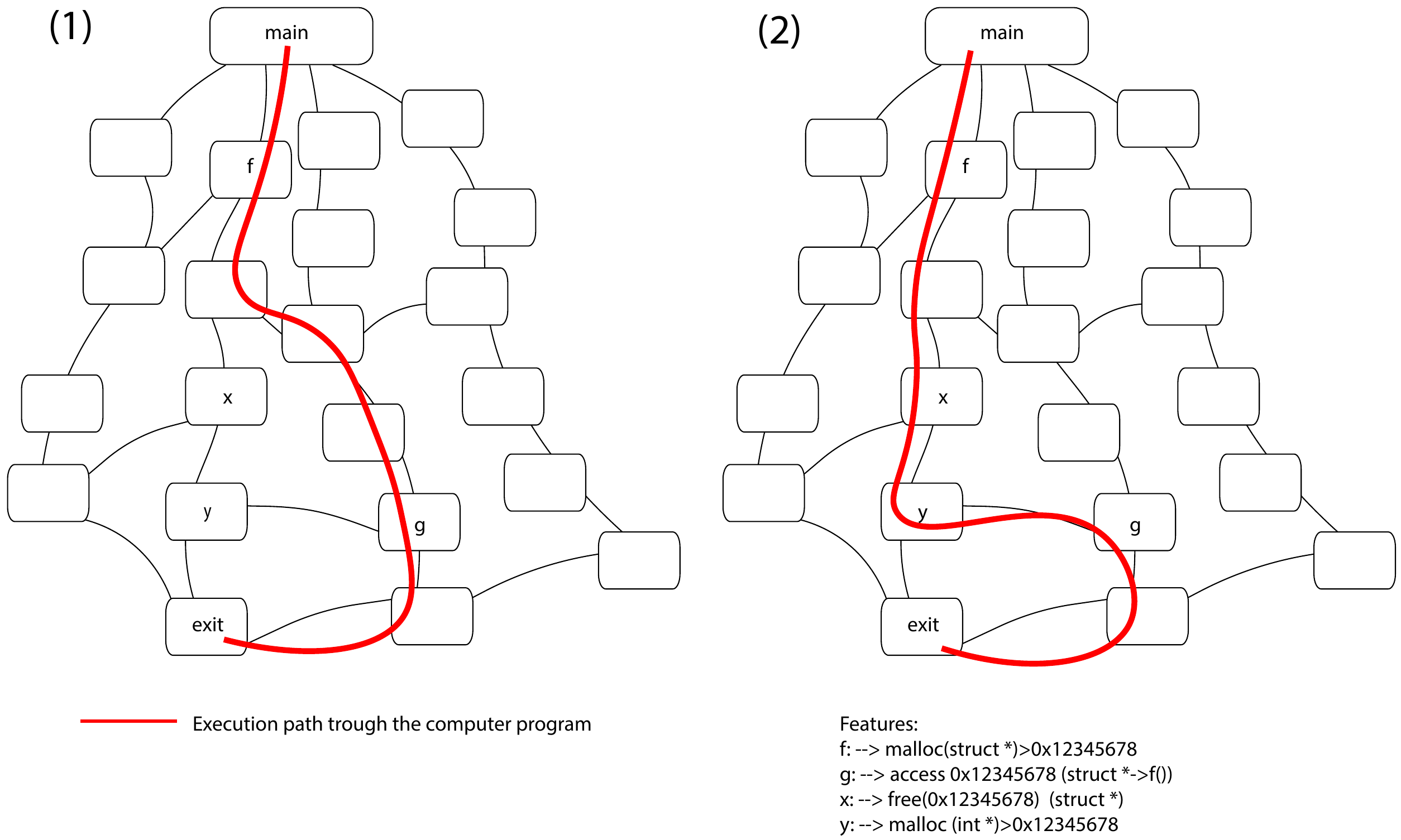}
    \caption{Detecting a Use-After-Free vulnerability based on traversing the CFG using extracted features}
    \label{fig:cfgtraversal}
\end{figure}

In Figure \ref{fig:cfgtraversal} we see a CFG of a computer program and two possible execution paths marked in red. Path (1) on the left side of the figure is the path the programmer originally intended. The path does not cause any problems, since the features of $f$ and $g$ are just \texttt{malloc} and \texttt{free} used as intended.
However, path (2) on the right side of the figure shows a path where first, memory is allocated, and later freed and reallocated with a different data type and used assuming the datatype from the first allocation. This is a classic Use-After-Free issue, which typically can lead to remote code execution. 
When tracing the paths and searching for matching weakness pattern a state-machine can be used. When the entrance state of a possible weakness is detected, the state machine starts tracing the weakness model, as long as it matches with the program flow. If an operation is executed that does not match the weakness model, the state machine is reset. However, if the weakness model is completed, a potential vulnerability is detected, which still needs to be verified.

We can describe a Use-After-Free vulnerability as an access to a memory area using the wrong datatype or accessing a previously free memory area. 
However, to detect a Use-After-Free vulnerability we need to symbolically simulate our heap and our state machine needs to start tracing at a free statement on a specific address. Next we trace if a path of the computer program can lead to allocating the same or an overlapping memory area again with a different data type. If this happens and an access to the memory area by using a previously allocated type (not matching the current state/data type of the heap-memory) is detected, we detected a possible Use-After-Free vulnerability.

In our case (1) in Figure \ref{fig:cfgtraversal} we detect a \texttt{malloc} when function $f$ is executed and we start tracing the memory area. However, there is only one access to the memory area when function $g$ is executed, but no free operation is happening. Thus, the path is considered to be safe, even though it clearly leaks memory (which could also be reported to the user).
However, in case (2) we have a \texttt{malloc} in $f$, a free in $x$ and a reallocation of the same memory area in function $y$. In function $g$ the memory area is accessed using the data type allocated in $f$, not matching the memory area, since it was freed and allocated with a different data type in $y$, therefore we raise a warning of a potentially Use-After-Free vulnerability. 

To verify if the vulnerability actually exists we use a form of guided symbolic execution. Since we know the call path which leads us to the situation, where the Use-After-Free potentially occurs, we can follow exactly its path. In our case, we execute the path backwards, since it makes it easier to verify if the path is possible. For this we use guided-backward symbolic execution. From function $g$, we check which constraints on the parameter of $g$ are required to call the access operation by solving the symbolic constraints. Next, we check if the previous function in the call graph can provide the parameter, and so on till we reach the user input. If a user input can be generated triggering the Use-After-Free situation, we confirmed the Use-After-Free vulnerability and the guided-backward symbolic execution can provide a sample input. This way, it supports the programmer to debug and fix the vulnerability. 

We use the guided-backward symbolic execution, since it is more likely a condition which is close to the potential vulnerability prevents the vulnerability from being exploitable. Therefore, if we figure out, that there is no possible input for $g$ to cause the access to the relevant memory area, we can directly abort the guided-backward symbolic execution and ruled out the possibility for a Use-After-Free vulnerability.

\paragraph{Guided-Backward-Symbolic Execution}~\cite{ma2011directed} describes a concept where we are at a location $l$ in a computer program, we know the execution path $p$, how it lead to the location $l$, and we want to compute input required for the computer program to execute the path $p$. 

\begin{figure}
    \centering
    \includegraphics[width=0.9\textwidth]{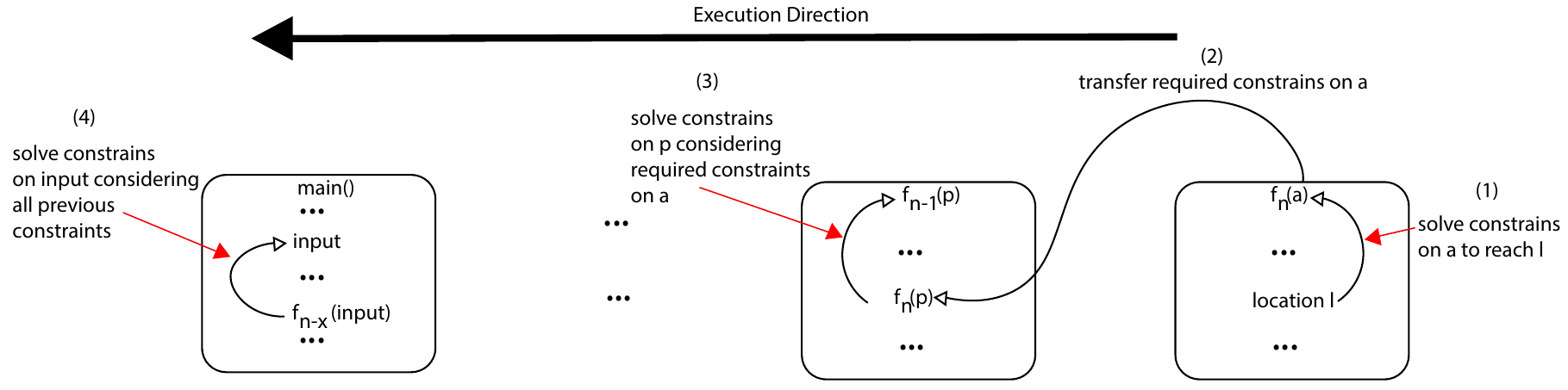}
    \caption{Guided-Backward Symbolic Execution}
    \label{fig:guidedBackwardSymex}
\end{figure}

As shown in Figure \ref{fig:guidedBackwardSymex}, we start execution at the location $l$ in the function $f_n$ and we solve under which constraints on the input parameter $a$ of $f_n$ the location $l$ can be reached. Next, we move to the function $f_{n-1}$ which is the directly predecessor of $f_n$ in the execution path ($f_{n-1}$ calls $f_n$) and we solve under which constraints for the input parameter $p$ of $f_{n-1}$ the required constraints of $a$ to reach $l$ are fulfilled. Afterwards we check to the predecessor of $f_{n-1}$ until either the constraints are not satisfiable or we reach user input. In case the constraints are not satisfiable, the location $l$ cannot be reached by the given path $p$, in case we reach user input, we can compute what the user needs to enter into the computer program to reach the location $l$ using the path $p$.


\section{Weakness Modeling}
\label{sec:vul}

To be able to check a CFG for weaknesses using our extracted feature, we need to define weakness models. As mentioned above a weakness model is a state machine which describes the concept of a specific weakness. Depending on the features we extracted, we can model and detect any weakness. However, in this paper we focus on certain weaknesses such as \emph{Use-After-Free} or \emph{Improper Neutralization of Special Elements}, which are both quite common weaknesses.

\subsection{Modeling Use-After Free}

To create a weakness model, we need to understand in detail how a certain weakness works. For a Use-After-Free Weakness this means that we have a memory area which is freed and afterward used by a different data type while there is still a dangling pointer accessing the memory area expecting a data type which is not stored there anymore. This can lead to pointers destinations or indirect calls being overwritten, and therefore corrupt the memory or the program flow of a computer program. 
We can break down a Use-After-Free vulnerability to the following actions:
\begin{enumerate}
    \item A previous allocated memory area of data type $d$ is freed. 
    \item New memory area of a different data type is allocated and overlaps the previously freed memory area. 
    \item A dangling pointer of data type $d$ is used to access the memory area
\end{enumerate}
For an Use-After-Free vulnerability, theoretical step (2) is not required, but without having new data stored in the memory area it is not possible to exploit the bug. 

For our detection system, this means, the state machine to model Use-After-Free vulnerabilities should look as the following:
\begin{verbatim}
    1. free(addr1)
    2. malloc(addr2) where addr2 == addr1 or 
             addr2 to addr2+size2 overlaps addr1 to addr1+size1
    3. pointer with type of addr1 is used to access
             overlapping memory area
\end{verbatim}
Both read or write operations through a dangling pointer can be very dangerous. 

\section{Theoretical Analysis}

In this section, we present a formal analysis of our divide-and-conquer symbolic execution approach, focusing on its soundness, completeness and complexity with respect to memory corruption vulnerability detection.

\subsection{Soundness Analysis}

We begin by defining soundness in the context of our vulnerability detection approach:

\begin{definition}[Soundness]
An approach to memory corruption vulnerability detection is sound if it does not miss any memory corruption vulnerabilities that would be found by exhaustive symbolic execution of the entire program.
\end{definition}

We now state and prove the soundness theorem for our approach:

\begin{theorem}[Soundness]
The divide-and-conquer symbolic execution approach with smart symbolic summaries is sound with respect to detecting memory corruption vulnerabilities.
\end{theorem}

\begin{proof}
We prove the soundness of our approach by showing that:
\begin{enumerate}[label=(\alph*)]
    \item Function summaries capture all relevant behaviors
    \item The recombination process preserves memory corruption vulnerability detection
\end{enumerate}

\textbf{Step 1: Function Summary Completeness}

Let $f$ be an arbitrary function in the program $P$, and let $\Sigma(f)$ be the symbolic summary of $f$.

\begin{lemma}
For any possible execution of $f$ with precondition $\alpha$, the corresponding postcondition $\omega$ and side effects $\theta$ related to memory operations are captured in $\Sigma(f)$.
\end{lemma}

\begin{proof}[Proof of Lemma 1]
Our symbolic execution of $f$ considers all possible paths through the function. For each path, we extract the full postcondition and all relevant side effects, including memory allocations, deallocations, and accesses. The summary $\Sigma(f)$ is constructed as the union of all these path results. Therefore, $\Sigma(f)$ captures all possible memory-related behaviors of $f$.
\end{proof}

\textbf{Step 2: Recombination Preservation}

Let $G$ be the control flow graph of program $P$.

\begin{lemma}
If a memory corruption vulnerability exists in $P$, there exists a path in $G$ where the vulnerability can be detected using the function summaries.
\end{lemma}

\begin{proof}[Proof of Lemma 2]
Our approach traverses all paths in $G$. At each function call, we apply the corresponding function summary. The side effects $\theta$ capture all memory operations, including allocations, deallocations, and accesses. Our vulnerability model checks for memory corruption by analyzing these extracted heap operations, including:
\begin{itemize}
    \item Buffer overflows
    \item Use-after-free vulnerabilities
    \item Double-free vulnerabilities
\end{itemize}
The guided backward symbolic execution verifies the feasibility of potential vulnerabilities. Therefore, if a memory corruption vulnerability exists, it will be captured in the side effects and detected by our vulnerability model.
\end{proof}

By Lemma 1 and Lemma 2, our approach does not miss any memory corruption vulnerabilities that would be detected by full symbolic execution. Thus, the approach is sound.
\end{proof}

\subsection*{Assumptions and Limitations}
While our soundness proof demonstrates the theoretical strength of our approach, it relies on certain assumptions:
\begin{itemize}
    \item The correctness of the underlying symbolic execution engine (angr in our implementation).
    \item The accuracy of our vulnerability model for memory corruption.
    \item The ability to accurately extract and represent all relevant memory-related side effects in function summaries.
\end{itemize}
It's important to note that while our approach is sound (does not miss vulnerabilities), it may produce false positives that require additional verification. This is a trade-off we make to improve performance and scalability.

\subsection{Completeness Analysis}

While soundness ensures that our approach doesn't miss any vulnerabilities, completeness addresses whether all reported vulnerabilities are genuine. In the context of our divide-and-conquer symbolic execution approach, we define completeness as follows:

\begin{definition}[Completeness]
An approach to memory corruption vulnerability detection is complete if every vulnerability it reports corresponds to a genuine vulnerability in the program under analysis.
\end{definition}

Our approach prioritizes soundness over completeness, which may lead to false positives. However, we can provide a degree of completeness through our verification step:

\begin{theorem}[Partial Completeness]
The divide-and-conquer symbolic execution approach with smart symbolic summaries and guided backward symbolic execution verification is partially complete with respect to detecting memory corruption vulnerabilities.
\end{theorem}

\begin{proof}
We demonstrate partial completeness through the following steps:

\begin{enumerate}
    \item Initial Detection: Our approach may flag potential vulnerabilities based on function summaries and their recombination.
    \item Verification: For each potential vulnerability, we perform guided backward symbolic execution to verify its feasibility.
    \item False Positive Elimination: If the backward symbolic execution cannot generate a concrete input triggering the vulnerability, we eliminate it as a false positive.
\end{enumerate}

Let $V$ be the set of all reported vulnerabilities after the verification step. For each $v \in V$:

\begin{itemize}
    \item There exists a feasible path $p$ in the program's control flow graph $G$ leading to $v$.
    \item There exists a concrete input $i$ that causes the program to follow path $p$ and trigger vulnerability $v$.
\end{itemize}

Therefore, each reported vulnerability corresponds to a genuine vulnerability in the program.
\end{proof}
However, our approach may still produce false positives, since our analysis might not account for all environmental constraints that would prevent a vulnerability from being exploited in practice.

\subsection{Complexity Analysis}

Let's analyze the complexity of our divide-and-conquer symbolic execution approach more rigorously. We'll consider the following parameters:
\begin{itemize}
    \item $n$: number of functions in the program
    \item $m$: average number of paths in a function
    \item $k$: maximum depth of the call graph
    \item $p$: average number of parameters per function
\end{itemize}

Our approach consists of two main phases: (1) generating function summaries and (2) analyzing the program using these summaries.

\subsubsection{Function Summary Generation}

For each function:
\begin{itemize}
    \item Time complexity: $O(m)$ for symbolic execution of all paths
    \item Space complexity: $O(mp)$ to store the summary (preconditions and postconditions for each path)
\end{itemize}

Total for all functions:
\begin{itemize}
    \item Time: $O(nm)$
    \item Space: $O(nmp)$
\end{itemize}

\subsubsection{Program Analysis Using Summaries}

In the worst case, we may need to consider all combinations of paths through the call graph:
\begin{itemize}
    \item Time complexity: $O(m^k)$, as we may need to explore $m$ options at each of the $k$ levels of the call graph
    \item Space complexity: $O(k)$ for the call stack
\end{itemize}

\subsubsection{Verification Step}

For each potential vulnerability:
\begin{itemize}
    \item Time complexity: $O(k)$ for backward traversal of the call graph
    \item Space complexity: $O(k)$ for storing the path conditions
\end{itemize}

Assuming we find $v$ potential vulnerabilities, the total for verification is:
\begin{itemize}
    \item Time: $O(vk)$
    \item Space: $O(k)$
\end{itemize}

\subsubsection{Overall Complexity}

\begin{itemize}
    \item Time: $O(nm + m^k + vk)$
    \item Space: $O(nmp + k)$
\end{itemize}

In comparison using full symbolic execution, the entire program is analyzed as a whole:
\begin{itemize}
    \item Time: $O(m^n)$ (worst case)
    \item Space: $O(m^n)$ (worst case)
\end{itemize}

Comparing these complexities to our approach:
\begin{itemize}
    \item Our Time: $O(nm + m^k + vk)$
    \item Our Space: $O(nmp + k)$
\end{itemize}

It's worth noting that our approach has theoretically the same worst-case time complexity as Godefroid's compositional symbolic execution \cite{Godefroid2007}, which also achieves $O(m^k)$ for the analysis phase. However, our approach offers a practical advantage by removing features that are not relevant to vulnerability detection, thus reducing the effective state space traversing the function summaries and searching for vulnerabilities.

To formalize this advantage, let's introduce a new parameter:
\begin{itemize}
    \item $r$: the average reduction factor in the number of paths due to feature removal $(0 < r \leq 1)$
\end{itemize}

With this parameter, we can express our effective time complexity as:
\begin{itemize}
    \item Effective Time: $O(nm + (rm)^k + vk)$
\end{itemize}

This formulation shows that our approach effectively reduces the branching factor from $m$ to $rm$ in the exponential term. In practice, $r$ can be significantly smaller than 1, leading to substantial performance improvements.

We can observe that:

\begin{enumerate}
    \item \textbf{Time Complexity:} Our approach's effective time complexity is generally better than both full symbolic execution and Godefroid's approach:
    \begin{itemize}
        \item The term $nm$ remains the same, growing linearly with the number of functions.
        \item $(rm)^k$ is smaller than $m^k$ when $r < 1$, which is typically the case due to our feature removal technique.
        \item The additional term $vk$ is usually negligible compared to $(rm)^k$ or $m^k$.
    \end{itemize}

    \item \textbf{Space Complexity:} Our approach offers a substantial improvement:
    \begin{itemize}
        \item $O(nmp + k)$ is typically much smaller than $O(m^n)$.
        \item Our space complexity grows linearly with the number of functions ($n$) and parameters ($p$), and only linearly with the call graph depth ($k$).
        \item The feature removal technique further reduces the effective space needed for each function summary.
    \end{itemize}
\end{enumerate}

It's important to note that these are worst-case complexities, and the actual value of $r$ will vary depending on the program being analyzed and the effectiveness of our feature removal technique. In practice, the performance difference can be even more pronounced, especially for large programs with many functions but a relatively shallow call graph. However, for programs consisting of a single long function or with a very deep call graph approaching the total number of functions, the performance of our approach may converge to that of full symbolic execution or Godefroid's approach.

\section{Evaluation}
\label{sec:eval}
For our evaluation, we implemented the divide algorithm using angr~\cite{shoshitaishvili2016state}. Angr is a symbolic execution and reverse engineering toolkit written in python. It performs symbolic execution on binaries, thus, no source code is required to run symbolic execution with angr. However, for our feature extraction, there may be situations where it is beneficial to have the source code available, for example, the feature extraction for Use-After-Free weaknesses requires exact type information. While in many cases the type can be reconstructed from the binary code, the results are more robust when the source code is available for providing exact type information. 

For source code analysis, we use \emph{llvm}~\cite{lattner2004llvm} to extract the same features from the source code as from the binary. This way, the information can be joined and enrich each other. Knowing the exact type information and the runtime malloc information from the symbolic execution, we can simulate the heap behavior then searching through the function call graph using our vulnerability model. The search is performed using a depth-first-search algorithm.

To understand the performance of our divide-and-conquer approach, we compare it with a classic symbolic execution approach, using angr, where we simulate the heap using the angr heap plugin to be able to detect Use-After-Free vulnerabilities. We run our evaluation on different test programs containing easier and hard to find Use-After-Free vulnerabilities. 
The test system had 128GB of memory and an AMD Ryzen 7950X CPU. However, while the amount of memory dramatically influences the problem size which can be solved, especially when not splitting the program into functions, the CPU only influences the speed, and, since our software is not optimized for multi-core yet, the CPU does not have a very big influence. However, we used a time limit of 10 minutes. As an operating system, we use Ubuntu 22.04.

\begin{figure}
    \centering
    \includegraphics[width=0.75\linewidth]{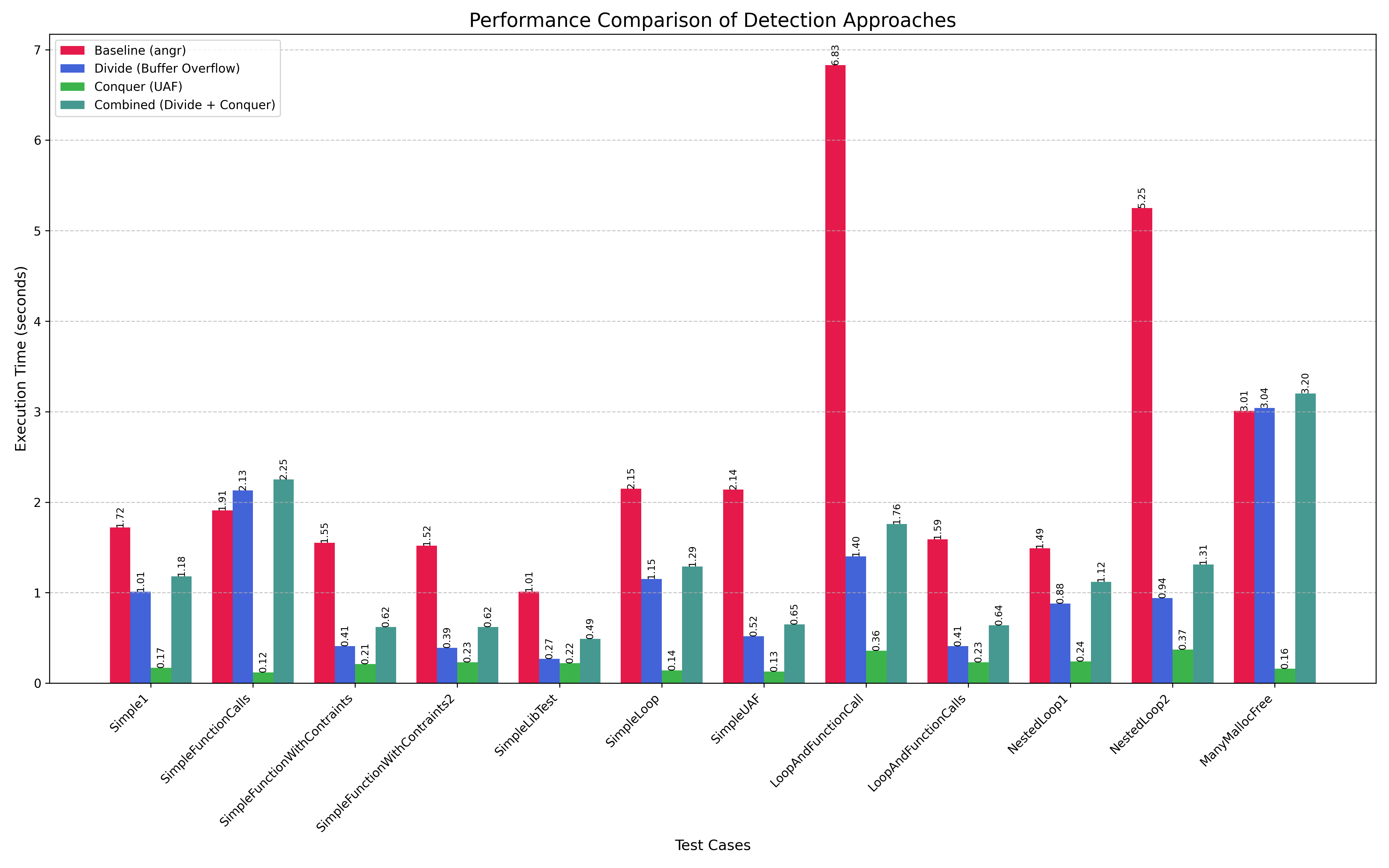}
    \caption{Comparison of the different analysis methods}
    \label{fig:performance1}
\end{figure}

For our evaluation, we have 12 programs with different vulnerabilities, from buffer overruns to formatted string errors, memory leaks, and Use-After-Free vulnerabilities. Some of the test programs have loops and other obstacles before the vulnerability can be reached. In our evaluation, we measure whether the reference implementation and our divide-and-conquer approach can find the vulnerability in each sample program and how much time it took to find the vulnerability. We do not use any swap memory, whenever we reach 100GB we kill the test and count the vulnerability as not found. For the 12 simple programs, the memory limit was not hit. 
The results of the first test is shown in Figure~\ref{fig:performance1}.

For small programs without many function calls it is visible that the divide process is not superior, while for larger programs with many calls the advantage is clearly visible.
In the conquer phase the guided-backward symbolic execution (to verify the potential Use-After-Free vulnerabilities) takes for all of the examples more time than the graph analysis. 

In a next test, we use more complex control structures and function call graphs. Therefore, we use programs with loops and recursion, such as binary search trees and binary search. These programs are traditionally challenging for symbolic execution. 
We check for how many of these programs our divide and conquer approach can find the vulnerability compared to the baseline implementation of angr. All these programs had a hidden Use-After-Free vulnerability, to ensure the divide and conquer approach needs to go full way. 

\begin{figure}
    \centering
    \includegraphics[width=0.70\linewidth]{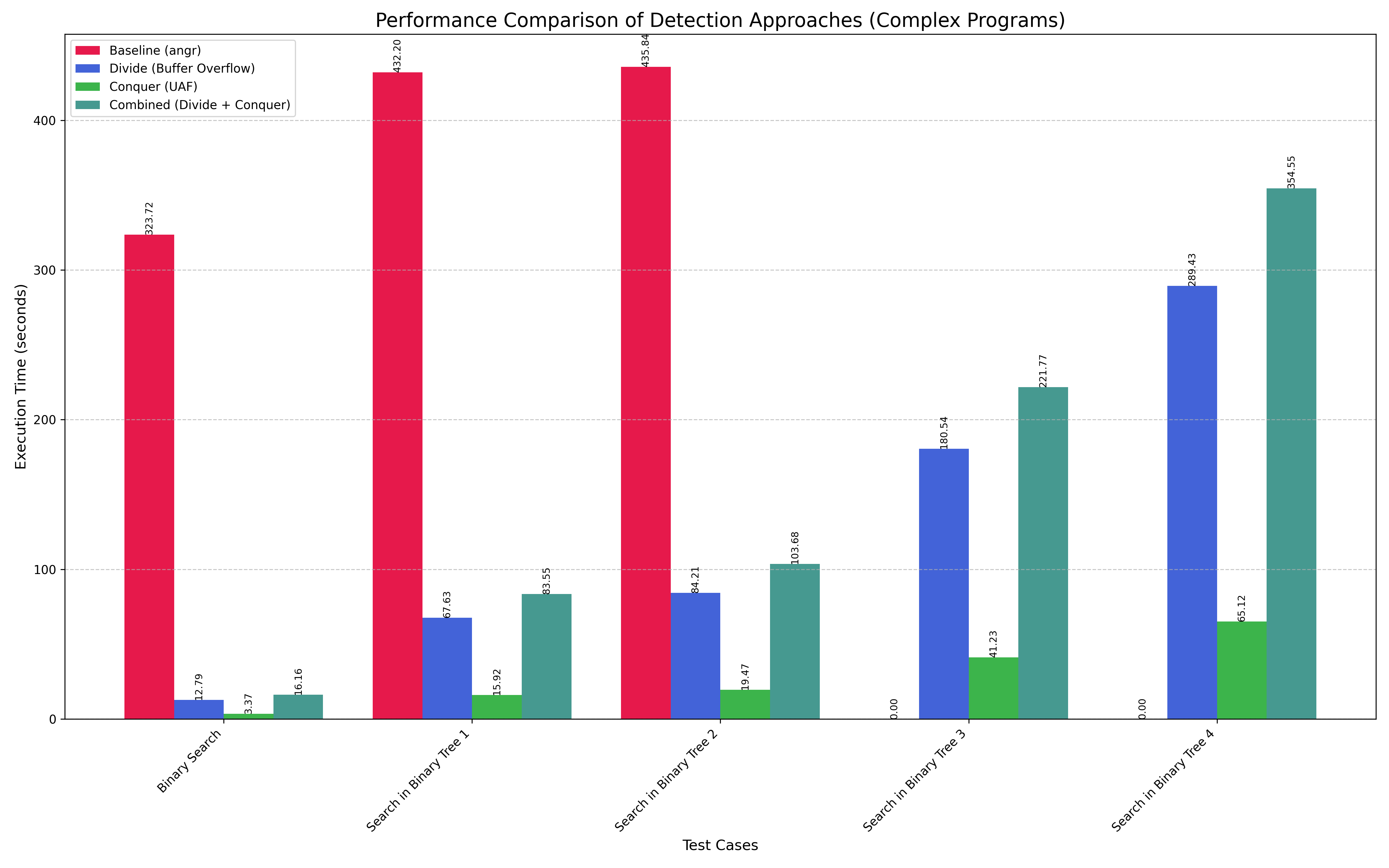}
    \caption{Comparison of the different analysis methods with complex programs}
    \label{fig:performance2}
\end{figure}

We see in Figure~\ref{fig:performance2}, that angr's base-line implementation reaches a limit where it runs out of memory. However, the divide and conquer approach can solve the given test programs. Adding longer loops into functions or also having too complex logic within a single function, the divide-and-conquer approach could fail, too. 
Last, we tested our approach with real software; however, for larger software also the divide and conquer approach often fails. However, for smaller programs used in specific context such as in industrial control systems, microcontroller or in lab-software, the approach is promising and can help to remove errors, bugs and vulnerabilities during the development phase. 

\section{Conclusion}
\label{sec:conclusion}
In this paper, we demonstrated a divide-and-conquer approach for symbolic execution, focusing specifically on vulnerability detection. Our method strategically ignores and strips away program parts that are not essential for identifying vulnerabilities, such as non-security-critical functions or redundant code paths. This targeted approach reduces the runtime of symbolic execution and vulnerability detection compared to traditional, full-program symbolic execution techniques. As a result, we can analyze larger codebases and potentially uncover more bugs.
While our approach is not yet efficient enough to verify the absence of vulnerabilities in large, real-world software systems, it does make such verification more feasible for smaller programs compared to unoptimized symbolic execution methods. This represents a step towards proving the absence of certain weakness classes in software. However, further optimization is required before this technique becomes broadly applicable outside of specialized domains.
Our method shows particular promise for critical infrastructure and other high-assurance systems where comprehensive security analysis is paramount. Nevertheless, challenges remain, particularly in handling complex control structures like loops and improving the scalability of symbolic execution for larger systems. Addressing these issues will be crucial areas for future research to fully realize the potential of this approach in practical software security verification.

%
%
%
 \bibliographystyle{splncs04}
 \bibliography{nordsecbib}

\begin{thebibliography}{10}
\providecommand{\url}[1]{\texttt{#1}}
\providecommand{\urlprefix}{URL }
\providecommand{\doi}[1]{https://doi.org/#1}

\bibitem{anand2008demand}
Anand, S., Godefroid, P., Tillmann, N.: Demand-driven compositional symbolic
  execution. In: Tools and Algorithms for the Construction and Analysis of
  Systems: 14th International Conference, TACAS 2008, Held as Part of the Joint
  European Conferences on Theory and Practice of Software, ETAPS 2008,
  Budapest, Hungary, March 29-April 6, 2008. Proceedings 14. pp. 367--381.
  Springer (2008)

\bibitem{ani2017review}
Ani, U.P.D., He, H., Tiwari, A.: Review of cybersecurity issues in industrial
  critical infrastructure: manufacturing in perspective. Journal of Cyber
  Security Technology  \textbf{1}(1),  32--74 (2017)

\bibitem{baldoni2018survey}
Baldoni, R., Coppa, E., D’elia, D.C., Demetrescu, C., Finocchi, I.: A survey
  of symbolic execution techniques. ACM Computing Surveys (CSUR)
  \textbf{51}(3),  1--39 (2018)

\bibitem{beyer2011cpachecker}
Beyer, D., Keremoglu, M.E.: Cpachecker: A tool for configurable software
  verification. In: Computer Aided Verification: 23rd International Conference,
  CAV 2011, Snowbird, UT, USA, July 14-20, 2011. Proceedings 23. pp. 184--190.
  Springer (2011)

\bibitem{chan1998model}
Chan, W., Anderson, R.J., Beame, P., Burns, S., Modugno, F., Notkin, D., Reese,
  J.D.: Model checking large software specifications. IEEE Transactions on
  software Engineering  \textbf{24}(7),  498--520 (1998)

\bibitem{chen2018systematic}
Chen, C., Cui, B., Ma, J., Wu, R., Guo, J., Liu, W.: A systematic review of
  fuzzing techniques. Computers \& Security  \textbf{75},  118--137 (2018)

\bibitem{chipounov2011s2e}
Chipounov, V., Kuznetsov, V., Candea, G.: S2e: A platform for in-vivo
  multi-path analysis of software systems. Acm Sigplan Notices  \textbf{46}(3),
   265--278 (2011)

\bibitem{clarke2004tool}
Clarke, E., Kroening, D., Lerda, F.: A tool for checking ansi-c programs. In:
  Tools and Algorithms for the Construction and Analysis of Systems: 10th
  International Conference, TACAS 2004, Held as Part of the Joint European
  Conferences on Theory and Practice of Software, ETAPS 2004, Barcelona, Spain,
  March 29-April 2, 2004. Proceedings 10. pp. 168--176. Springer (2004)

\bibitem{cook2013ranking}
Cook, B., Kroening, D., R{\"u}mmer, P., Wintersteiger, C.M.: Ranking function
  synthesis for bit-vector relations. Formal methods in system design
  \textbf{43}(1),  93--120 (2013)

\bibitem{godefroid2007compositional}
Godefroid, P.: Compositional dynamic test generation. In: Proceedings of the
  34th annual ACM SIGPLAN-SIGACT symposium on Principles of programming
  languages. pp. 47--54 (2007)

\bibitem{haltermann2024ranged}
Haltermann, J., Jakobs, M.C., Richter, C., Wehrheim, H.: Ranged program
  analysis: A parallel divide-and-conquer approach for software verification.
  In: Software Engineering 2024 (SE 2024). pp. 157--158. Gesellschaft f{\"u}r
  Informatik eV (2024)

\bibitem{havelund2000model}
Havelund, K., Pressburger, T.: Model checking java programs using java
  pathfinder. International Journal on Software Tools for Technology Transfer
  \textbf{2},  366--381 (2000)

\bibitem{hensel2018termination}
Hensel, J., Giesl, J., Frohn, F., Str{\"o}der, T.: Termination and complexity
  analysis for programs with bitvector arithmetic by symbolic execution.
  Journal of Logical and Algebraic Methods in Programming  \textbf{97},
  105--130 (2018)

\bibitem{imtiaz2021comparative}
Imtiaz, N., Thorn, S., Williams, L.: A comparative study of vulnerability
  reporting by software composition analysis tools. In: Proceedings of the 15th
  ACM/IEEE International Symposium on Empirical Software Engineering and
  Measurement (ESEM). pp. 1--11 (2021)

\bibitem{s23239407}
Kim, J., Park, J.: Enhancing security of web-based iot services via xss
  vulnerability detection. Sensors  \textbf{23}(23) (2023).
  \doi{10.3390/s23239407}, \url{https://www.mdpi.com/1424-8220/23/23/9407}

\bibitem{king1976symbolic}
King, J.C.: Symbolic execution and program testing. Communications of the ACM
  \textbf{19}(7),  385--394 (1976)

\bibitem{lattner2004llvm}
Lattner, C., Adve, V.: Llvm: A compilation framework for lifelong program
  analysis \& transformation. In: International symposium on code generation
  and optimization, 2004. CGO 2004. pp. 75--86. IEEE (2004)

\bibitem{li2013software}
Li, H., Kim, T., Bat-Erdene, M., Lee, H.: Software vulnerability detection
  using backward trace analysis and symbolic execution. In: 2013 International
  Conference on Availability, Reliability and Security. pp. 446--454. IEEE
  (2013)

\bibitem{lin2015compositional}
Lin, Y., Miller, T., S{\o}ndergaard, H.: Compositional symbolic execution using
  fine-grained summaries. In: 2015 24th Australasian Software Engineering
  Conference. pp. 213--222. IEEE (2015)

\bibitem{luckow2020complexity}
Luckow, K., Kersten, R., Pasareanu, C.: Complexity vulnerability analysis using
  symbolic execution. Software Testing, Verification and Reliability
  \textbf{30}(7-8),  e1716 (2020)

\bibitem{ma2011directed}
Ma, K.K., Yit~Phang, K., Foster, J.S., Hicks, M.: Directed symbolic execution.
  In: Static Analysis: 18th International Symposium, SAS 2011, Venice, Italy,
  September 14-16, 2011. Proceedings 18. pp. 95--111. Springer (2011)

\bibitem{malmain2024libafl}
Malmain, R., Fioraldi, A., Aur{\'e}lien, F.: Libafl qemu: A library for
  fuzzing-oriented emulation. In: BAR 2024, Workshop on Binary Analysis
  Research, colocated with NDSS 2024 (2024)

\bibitem{MITRECWE}
{MITRE Corporation}: Common weakness enumeration (cwe). Tech. rep., {MITRE
  Corporation} (2024), \url{https://cwe.mitre.org/}

\bibitem{NIST800-30}
{National Institute of Standards and Technology}: Nist special publication
  800-30: Risk management guide for information technology systems. Tech. rep.,
  {National Institute of Standards and Technology} (2002),
  \url{https://nvlpubs.nist.gov/nistpubs/Legacy/SP/nistspecialpublication800-30.pdf}

\bibitem{NIST2012}
{National Institute of Standards and Technology}: Nist special publication
  800-30 revision 1: Guide for conducting risk assessments. Tech. rep.,
  {National Institute of Standards and Technology} (2012),
  \url{https://nvlpubs.nist.gov/nistpubs/Legacy/SP/nistspecialpublication800-30r1.pdf}

\bibitem{newsome2005dynamic}
Newsome, J., Song, D.X.: Dynamic taint analysis for automatic detection,
  analysis, and signaturegeneration of exploits on commodity software. In:
  NDSS. vol.~5, pp.~3--4. Citeseer (2005)

\bibitem{poeplau2020symbolic}
Poeplau, S., Francillon, A.: Symbolic execution with $\{$SymCC$\}$: Don't
  interpret, compile! In: 29th USENIX Security Symposium (USENIX Security 20).
  pp. 181--198 (2020)

\bibitem{ramos2015under}
Ramos, D.A., Engler, D.: $\{$Under-Constrained$\}$ symbolic execution:
  Correctness checking for real code. In: 24th USENIX Security Symposium
  (USENIX Security 15). pp. 49--64 (2015)

\bibitem{schwartz2010all}
Schwartz, E.J., Avgerinos, T., Brumley, D.: All you ever wanted to know about
  dynamic taint analysis and forward symbolic execution (but might have been
  afraid to ask). In: 2010 IEEE symposium on Security and privacy. pp.
  317--331. IEEE (2010)

\bibitem{serebryany2012addresssanitizer}
Serebryany, K., Bruening, D., Potapenko, A., Vyukov, D.:
  $\{$AddressSanitizer$\}$: A fast address sanity checker. In: 2012 USENIX
  annual technical conference (USENIX ATC 12). pp. 309--318 (2012)

\bibitem{sharma2012interpolants}
Sharma, R., Nori, A.V., Aiken, A.: Interpolants as classifiers. In:
  International Conference on Computer Aided Verification. pp. 71--87. Springer
  (2012)

\bibitem{shoshitaishvili2016state}
Shoshitaishvili, Y., Wang, R., Salls, C., Stephens, N., Polino, M., Dutcher,
  A., Grosen, J., Feng, S., Hauser, C., Kruegel, C., Vigna, G.: {SoK: (State
  of) The Art of War: Offensive Techniques in Binary Analysis}. In: IEEE
  Symposium on Security and Privacy (2016)

\bibitem{Shostack2014}
Shostack, A.: Threat Modeling: Designing for Security. Wiley, Indianapolis, IN
  (2014)

\bibitem{stephens2016driller}
Stephens, N., Grosen, J., Salls, C., Dutcher, A., Wang, R., Corbetta, J.,
  Shoshitaishvili, Y., Kruegel, C., Vigna, G.: Driller: Augmenting fuzzing
  through selective symbolic execution. In: NDSS. vol.~16, pp. 1--16 (2016)

\bibitem{tang2022accelerating}
Tang, S., Wang, X., Gao, Y., Hu, W.: Accelerating soc security verification and
  vulnerability detection through symbolic execution. In: 2022 19th
  International SoC Design Conference (ISOCC). pp. 207--208. IEEE (2022)

\bibitem{trabish2018chopped}
Trabish, D., Mattavelli, A., Rinetzky, N., Cadar, C.: Chopped symbolic
  execution. In: Proceedings of the 40th International Conference on Software
  Engineering. pp. 350--360 (2018)

\bibitem{traxler2008time}
Traxler, P.: The time complexity of constraint satisfaction. In: Parameterized
  and Exact Computation: Third International Workshop, IWPEC 2008, Victoria,
  Canada, May 14-16, 2008. Proceedings 3. pp. 190--201. Springer (2008)

\bibitem{tu2023boosting}
Tu, H.: Boosting symbolic execution for heap-based vulnerability detection and
  exploit generation. In: 2023 IEEE/ACM 45th International Conference on
  Software Engineering: Companion Proceedings (ICSE-Companion). pp. 218--220.
  IEEE (2023)

\bibitem{van2012memory}
Van~der Veen, V., Dutt-Sharma, N., Cavallaro, L., Bos, H.: Memory errors: The
  past, the present, and the future. In: Research in Attacks, Intrusions, and
  Defenses: 15th International Symposium, RAID 2012, Amsterdam, The
  Netherlands, September 12-14, 2012. Proceedings 15. pp. 86--106. Springer
  (2012)

\bibitem{xie2015s}
Xie, X., Liu, Y., Le, W., Li, X., Chen, H.: S-looper: Automatic summarization
  for multipath string loops. In: Proceedings of the 2015 International
  Symposium on Software Testing and Analysis. pp. 188--198 (2015)

\bibitem{yao2017statsym}
Yao, F., Li, Y., Chen, Y., Xue, H., Lan, T., Venkataramani, G.: Statsym:
  vulnerable path discovery through statistics-guided symbolic execution. In:
  2017 47th Annual IEEE/IFIP International Conference on Dependable Systems and
  Networks (DSN). pp. 109--120. IEEE (2017)

\bibitem{yi2017eliminating}
Yi, Q., Yang, Z., Guo, S., Wang, C., Liu, J., Zhao, C.: Eliminating path
  redundancy via postconditioned symbolic execution. IEEE Transactions on
  Software Engineering  \textbf{44}(1),  25--43 (2017)

\bibitem{zhioua2014static}
Zhioua, Z., Short, S., Roudier, Y.: Static code analysis for software security
  verification: Problems and approaches. In: 2014 IEEE 38th International
  Computer Software and Applications Conference Workshops. pp. 102--109. IEEE
  (2014)

\bibitem{zhu2022fuzzing}
Zhu, X., Wen, S., Camtepe, S., Xiang, Y.: Fuzzing: a survey for roadmap. ACM
  Computing Surveys (CSUR)  \textbf{54}(11s),  1--36 (2022)

\end{thebibliography}

\end{document}